\documentstyle[12pt,aaspp4]{article}
\input epsf
\def\beq{\begin{equation}}
\def\eeq{\end{equation}}
\def\bey{\begin{eqnarray}}
\def\eey{\end{eqnarray}}

\newbox\grsign \setbox\grsign=\hbox{$>$} \newdimen\grdimen \grdimen=\ht\grsign
\newbox\simlessbox \newbox\simgreatbox
\setbox\simgreatbox=\hbox{\raise.5ex\hbox{$>$}\llap
     {\lower.5ex\hbox{$\sim$}}}\ht1=\grdimen\dp1=0pt
\setbox\simlessbox=\hbox{\raise.5ex\hbox{$<$}\llap
     {\lower.5ex\hbox{$\sim$}}}\ht2=\grdimen\dp2=0pt
\newcommand{\simgt}{\mathrel{\copy\simgreatbox}}

\newsavebox{\savepar}

\begin{document}

\title{Tidal Streams as Probes of the Galactic Potential}
\author{Kathryn V. Johnston\altaffilmark{1}, HongSheng
Zhao\altaffilmark{2}, David N. Spergel\altaffilmark{3} and Lars
Hernquist\altaffilmark{4}}

\altaffiltext{1}{Institute for Advanced Study, Princeton, NJ 08450}

\altaffiltext{2}{Sterrewacht Leiden, Niels Bohrweg 2, 2333 CA, Leiden,
The Netherlands}

\altaffiltext{3}{Princeton University Observatory, Princeton
University, Princeton, NJ 08544}

\altaffiltext{4}{Department of Astronomy and Astrophysics, University
of California, Santa Cruz, CA 95064.}

\abstract

We explore the use of tidal streams from Galactic satellites to
recover the potential of the Milky Way. Our study is motivated both by
the discovery of the first lengthy {\it stellar} stream in the halo
(\cite{it98}) and by the prospect of measuring proper motions of stars
brighter than 20th magnitude in such a stream with an accuracy of
$\sim 4\mu as/$yr, as will be possible with the Space Interferometry
Mission (SIM).  We assume that the heliocentric radial velocities of
these stars can be determined from supporting ground-based
spectroscopic surveys, and that the mass and phase-space coordinates
of the Galactic satellite with which they are associated will also
be known to SIM accuracy.
Using results from numerical simulations as trial data sets,
we find that, if we assume the correct form for the Galactic potential,
we can predict the distances to the stars as a
consequence of the narrow distribution of energy expected
along the streams.  
We develop an algorithm to
evaluate the accuracy of any adopted potential by requiring that
the satellite and stars recombine within a Galactic
lifetime when their current phase-space coordinates
are integrated backwards.
When applied to a four-dimensional
grid of triaxial logarithmic potentials, with varying circular
velocities, axis ratios and orientation of the major-axis
in the disk plane, the algorithm can recover the 
parameters used for the Milky Way
in a simulated data set to within a few percent
using only 100 stars in a tidal stream.

\keywords{Galaxy: fundamental parameters, halo, kinematics and dynamics,
structure}

\section{Introduction}

Tidal streams in the Galactic halo are a natural by-product of
hierarchical structure formation, in which galaxies build up their
mass by accreting less massive satellites.  These features can be
produced when matter, {\it i.e.} stars and/or gas, is liberated from a
companion either by tidal shocking at the pericenter of the
satellite's orbit or, in the case of globular clusters, through the
evaporation of stars across the tidal boundary imposed by the Milky
Way.  The stripped material populates leading and trailing tidal
streams that align with the orbit of the satellite for timescales
comparable to or greater than the age of the Galaxy
(e.g. \cite{ola95}; \cite{g95}; \cite{jsh95}; Johnston, Hernquist \&
Bolte 1996, hereafter \cite{jhb96}).

The notion of using tidal streams as Galactic potentiometers has been
considered previously by a number of authors (\cite{lb82}; \cite{k93};
\cite{g98}; \cite{z98}).  By assuming that several of the dwarf spheroidal
satellites are tidal debris, Lynden-Bell (1982) was able to obtain an
estimate of the mass of the Milky Way.  Similarly, if the Magellanic
Stream consists of gas tidally stripped from the Large and Small
Magellanic Clouds, it can be used to constrain the potential
(\cite{mf80}; \cite{llb82}; \cite{ljk95}).  
While these results are not definitive
measurements owing to the controversial nature of the Magellanic
Stream (\cite{md94}), they demonstrate the power of this approach.

There is now growing observational evidence for the existence of {\it
stellar} tidal streams in the halo.  Several globular clusters are
known to possess excess unbound stars outside their tidal radii
(\cite{g95}; \cite{ih95}; \cite{ksh96}) and there are also moving
groups in the halo with no obvious bound counterparts (for a review
see \cite{mhm96}).  On larger scales, the debris associated with the
dwarf galaxy recently discovered in the constellation Sagittarius
(\cite{igi94}), has now been identified in horizontal branch (HB) and
giant branch (GB) stars over $22^o$ along a Galactic great circle
roughly coinciding with the $l=0^o$ plane (\cite{m96}; \cite{a96};
\cite{fmrts96}; \cite{i97}). Moreover, Irwin \& Totten's (1998)
discovery of a carbon star trail encircling the Galaxy and aligned
with the same plane provides the first example of a data set that
samples the entire length of a stellar tidal stream.

Upcoming satellite missions promise to provide accurate phase-space
coordinates for individual stars in tidal streams.  NASA's Space
Interferometric Mission (SIM), scheduled for 2006, is a pointed
instrument that will detect stars as faint as 20th magnitude with
accuracies of a few $\mu as$.  ESA's Global Astrometric Interferometer
for Astrophysics (GAIA) will survey more than a billion stars across
the entire sky with an astrometric precision of $\le 10\mu as$.  This
represents an improvement over results obtained with the HIPPARCOS
satellite by a factor of about a thousand in accuracy and more than a
million in the volume sampled, and makes the idea of measuring the
proper motions of individual HB and GB stars (and hence carbon stars)
at distances $\le $ 100 kpc feasible.  At distances of tens of
kiloparsecs, these future missions, when supplemented with radial
velocities, will accurately measure five out of the six phase-space
coordinates of a star.

In this {\it Letter}, we investigate the extent to which the potential
of the Galaxy can be recovered using a data set such as Irwin \&
Totten's (1998) carbon star stream, assuming that phase-space
positions can be inferred with the accuracy of the SIM satellite.  At
the appropriate distances, parallaxes measured by SIM ($\sim 4\mu as$)
will be the least well-determined phase-space component.  The implied
uncertainty in distances, $\sim (D/20 {\rm kpc})^{2}$ kpc for stars at
a distance $D$, will exceed the thickness of the tidal stream.  In
\S3.1, we show that using energy conservation in the correct
background potential, together with the
other five phase-space coordinates, yields a more precise distance
estimate.  We describe simulations that provide trial data sets for
our proposed methods in \S 2, and use them to estimate the accuracy
of distances to debris stars constrained by energy conservation
in \S 3.1.  We apply our algorithm for recovering the
potential of the Milky Way in \S 3.2, and summarize conclusions in \S 4.

\section{Trial Data Sets and Galactic Model}

This paper is based on an analysis of simulated debris trails produced
by tidal disruption, as reported in \cite{jhb96}.  In these
simulations, the Milky Way potential $\Phi_{\rm MW}$
was taken to be smooth, static and
axisymmetric, and was represented by a three-component bulge-disk-halo
model. 
In each simulation, the satellite was represented by $10^4$ particles,
whose self-gravity was calculated using a self-consistent field code
(\cite{ho92}). The particles were initially distributed as a Plummer
model, and evolved in the Milky Way potential for 10 Gyrs. The reader
is referred to \cite{jhb96} for full details of the simulations.

We tested our algorithm on all the models (1-12) in JHB and, in the
figures which follow, we illustrate the results using their Model 11.
This satellite had an initial mass $2.893\times 10^7 M_{\odot}$ and
scale-length 0.602kpc.  It was on an orbit inclined at $50^{\circ}$
relative to the Galactic disk with an orbital period $\sim 2$Gyrs and
peri- and apo-center distances of $\sim 40$kpc and $\sim 160$kpc.  We
also performed two new simulations with the same initial conditions as in
Model 11, but with the spherical halo component of the Milky Way
potential replaced by 
oblate and triaxial components of the form
\beq
\Phi _{{\rm halo}}(x,y,z)={\frac{v_{{\rm circ}}^2}2}\ln (x^2+y^2/p^2+z^2/q^2+c^2)
\label{phalo}.
\eeq
In the remainder of
the paper we shall refer to these new simulations as Models 13
and 14.
In all cases $v_{circ}=181$km/s and $c=12$kpc.
In Models 1-12 $(p,q)=(1,1)$, in Model 13 $(p,q)=(1,0.75)$
and in Model 14  $(p,q)=(0.95,0.75)$.

\section{Results}

\subsection{Energy Distances}

\begin{figure}
\begin{center}
\epsscale{1.0}
\plotone{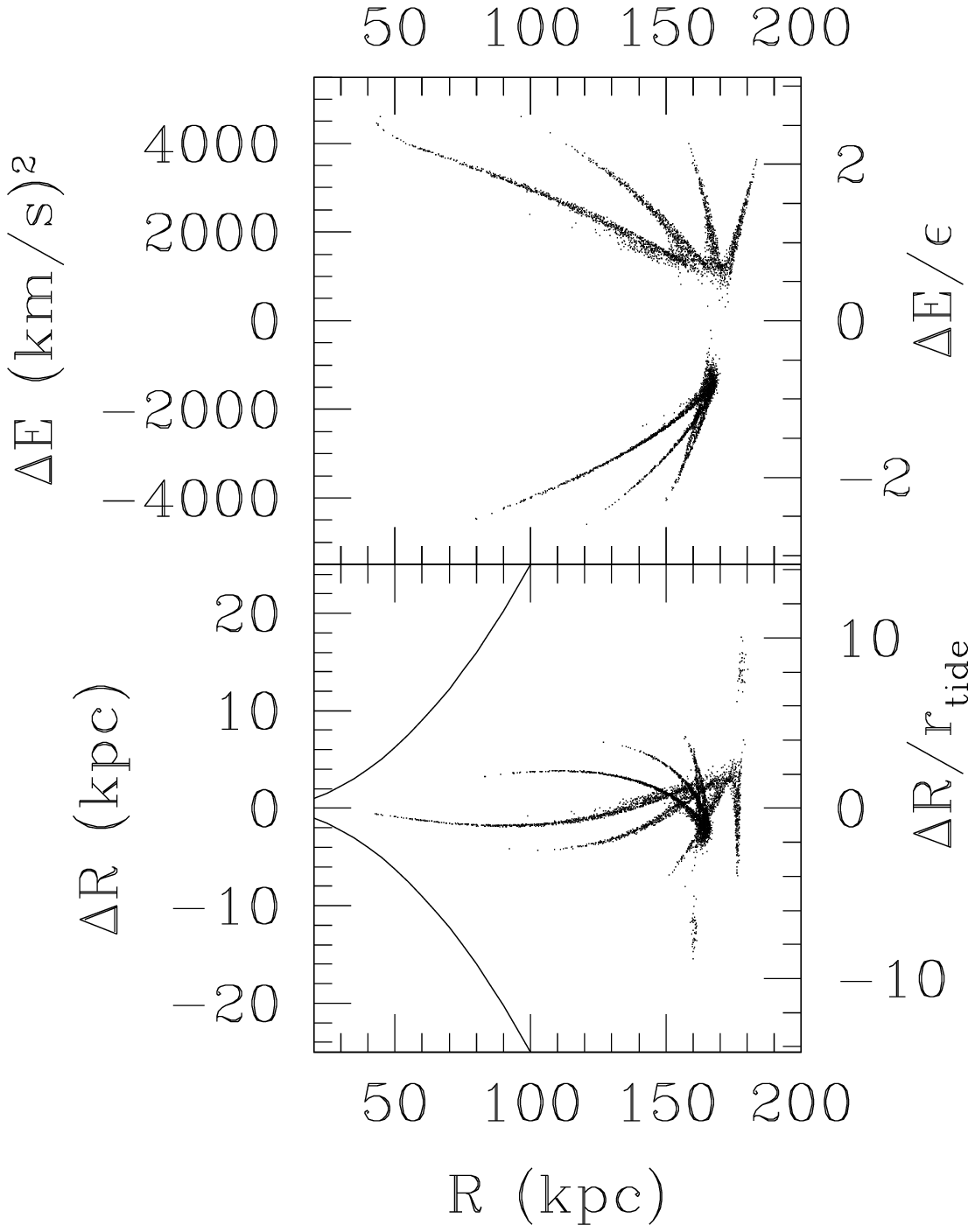}
\caption{Top panel --- orbital energy relative to the satellite's
$\Delta E$ for all particles unbound
at the end of the simulation of Model 11.
Bottom panel --- error in the distance estimate described in \S3.1.
The solid lines show the accuracy of distances derived from 
SIM parallax measurements.
\label{defig}}
\end{center}
\end{figure}

Tidal debris tends to become unbound from a satellite of mass $
m_{{\rm sat}}$ located at distance $R$ from the center of the Galaxy
on the physical scale given by the tidal radius
\begin{equation}  \label{rtide}
r_{{\rm tide}}=R \left( {\frac{m_{{\rm sat}} }{M_R}}\right)^{\frac{1 }{3}}
\end{equation}
(\cite{k62}), where $M_R$ is the mass of the Galaxy enclosed within
$R$.  Hence the orbital 
energies $E$ of material in the debris will lie approximately
within 
\begin{equation}
\epsilon=r_{{\rm tide}}{\frac{d\Phi }{dR}}=r_{{\rm tide}}{\frac{GM_R}{R^2}}
\label{de}
\end{equation}
of the satellite's orbital energy, $E_{\rm sat}$
(\cite{t93,j98}).
Equations (\ref{rtide}) and (\ref{de})
should be evaluated not at the current position of the
satellite, but at the pericenter of 
its orbit since most of the mass loss will occur
where the tidal field of the Milky Way is strongest.
The top panel
of Figure \ref{defig} shows $E$ for all particles which were no
longer bound to the satellite at the end of the simulation of Model
11, with the left hand axis in physical units and the right hand axis
in units relative to $E_{{\rm sat}}$ and scaled by $\epsilon$
given in equation (\ref{de}).  Note that the debris divides into two
populations, corresponding to stars in the leading/trailing stream
having $E$ offset by $\mp 5\epsilon/4$
from $E_{\rm sat}$. Each population has a width
$\sim 3 \epsilon/4$ about these average offsets.

Even with the astrometric accuracy of SIM, we will be able to measure
distances only to a precision $\simgt $10\% beyond 20kpc.  However, if
we can identify a population of stars which were once associated with
a satellite and we assume a form $\Phi_{\rm MW}$
for the Galactic potential, it is
possible to estimate their distances from the expected
energy distribution in the satellite's debris.
To test this idea
we ``observe'' the angular position $(l,b)$,
line of sight velocity $v_{{\rm los }}$ and proper motion
$(\mu_l,\mu_b)$ of each of the unbound particles at the
end of our simulations from an assumed
heliocentric viewpoint in the disk plane 8.5 kpc from the center of
the Galaxy, where the motions are defined in a Galactic rest frame.  
Each particle is assigned $E=E_{\rm sat}\mp 5\epsilon/4$
if it is ahead/behind the
satellite along the orbit,
and its heliocentric distance $d$ is then calculated by solving the
equation
\beq
	E={1\over 2} \left[v_{\rm los }^2+d^2(\mu_l^2+\mu_b^2)\right]
+\Phi_{\rm MW}(d,l,b).
\eeq
The bottom panel of Figure \ref{defig}
plots the error in the distance estimated for particles
in Model 11 using this method.
As can be seen from the right hand axis of this figure, we
expect this distance estimate to be good to within a few $r_{{\rm
tide}}$. In contrast, the solid lines show the much lower
accuracy of distances measured
from parallaxes with the SIM satellite (as described in \S 1).

\subsection{Constraining the Galactic Potential}

If we observe a set of stars that were once
associated with a satellite of known mass and phase-space
coordinates, then they must be on orbits
whose trajectories intersect with the satellite within the lifetime of
the Galaxy.  The following algorithm
assigns a quality factor for each assumed form $\Phi
(x,y,z)$ for the Galactic potential based on how well it can extrapolate
the orbits of tidal debris stars back into the progenitor satellite
galaxy.

\begin{enumerate}
\item  For each assumed potential we
integrate the satellite's orbit backwards and
calculate $r_{{\rm tide}}$ (eqn [\ref{rtide}]) and the energy
scale $\epsilon$ (eqn [\ref{de}]) at pericenter.

\item For each star in the debris with measured proper motion, angular
position and radial velocity we: (i) create $n_{{\rm test}}$ particles
with energies $E$ evenly distributed in the range $\pm 3\epsilon/4$ about
$(E_{{\rm sat }}\mp 5\epsilon/4)$ if
the star is ahead/behind the satellite along its orbit; (ii)
estimate the ``energy-distance'' to each particle; (iii)
integrate both the satellite and these particles backwards in time in
the assumed potential for a Galactic lifetime; and (iv) credit the
potential with a capture whenever any one of these particles is
separated from the satellite by a distance $dr<1.8\,r_{ {\rm tide}}$
and a velocity $dv<\sqrt{\frac{Gm_{{\rm sat}}}{dr}}$.

\item Assign the potential's ``score'' as the number of successful
captures, with the most likely potential having the highest score.
\end{enumerate}

\begin{figure}
\begin{center}
\epsscale{0.8}
\plotone{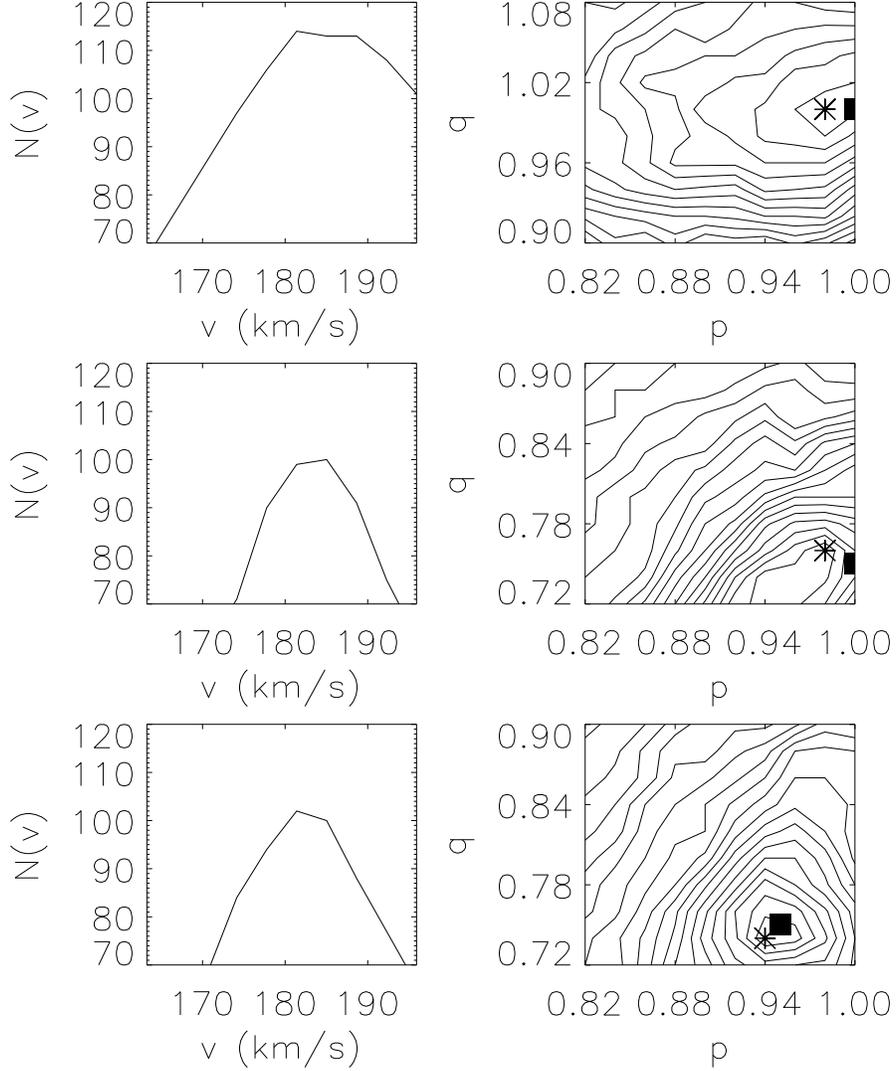}
\caption{Left hand panels ---
maximum number of captured particles for fixed
$v_{\rm circ}$ and arbitrary $q, p$ and axis orientation
as a result of applying our algorithm to Models 11
(top panel), 13 (middle panel) and 14 (bottom panel).
Right hand panels --- maximum number of rebound particles
contoured in the $p-q$ plane for the most-likely value
of $v_{\rm circ}$ identified in the left-hand panels.
The maxima corresponding to the most likely parameters
are marked with stars, and the parameters actually used 
in the simulations are shown with solid squares.
The contours are spaced by 4 particles. 
\label{algofig}}
\end{center}
\end{figure}

We tested this algorithm by applying it to 128 stars selected at
random at the end of each simulation.  For each star, 10 test
particles were integrated in a four-dimensional
grid of halo potentials of the form
given in equation (\ref{phalo}) with $v_{\rm circ}, q$ and $p$ varied by
10\% around the values in the simulations, and 
$x$ and $y$ axes rotated between 0-90 degrees.
The left hand panels of 
Figure \ref{algofig} show the maximum number of recaptured particles after
10 Gyrs
at fixed $v_{\rm circ}$ but varying $p, q$ and orientation of
the $x$ and $y$ axes, for Models 11 (top), 13 (middle) and
14 (bottom).
The right hand panels show contours of the maximum 
number of recaptured particles 
in the $p-q$ plane for arbitrary axis orientation
and $v_{\rm circ}$ corresponding to the maximum
in the left hand panels.
The points corresponding to the parameters of the potential in
which the simulations were originally run are marked with solid squares
and the recovered parameters are marked with stars. 
The figure demonstrates that this method is sensitive
to both the mass distribution and the geometry of the Milky Way.

\begin{figure}
\begin{center}
\epsscale{0.5}
\plotone{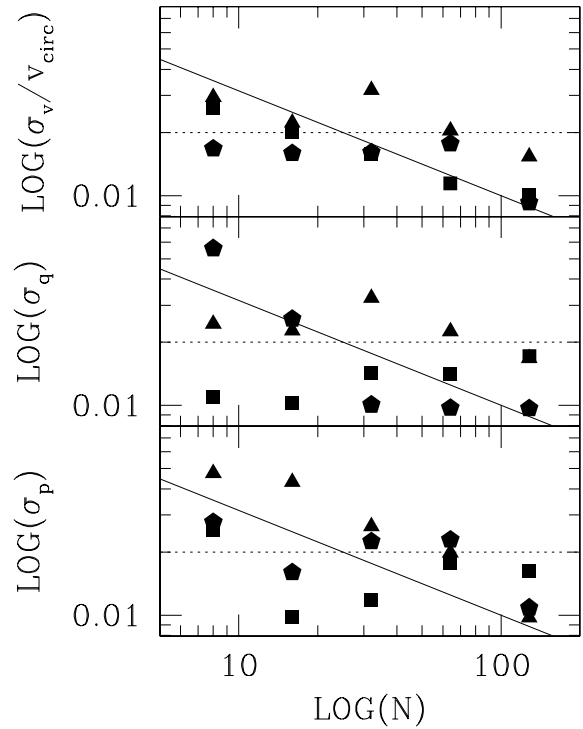}
\caption{Bootstrapped errors in the potential calculated with 
$N$ stars.  The solid triangles, squares and pentagons are for Models 11, 13
and 14 respectively. 
The bold line is given by $\sigma_w= 1/10\sqrt{N}$. The dotted 
line shows the size of one cell of the gridded distribution 
from which $\sigma_w$ was calculated.
\label{bootfig}}
\end{center}
\end{figure}

We can estimate the error introduced by using $N$ stars to determine
the potential from the bootstrap method.  We create $N(\log N)^2$
data sets, each of size $N$, by drawing stars at random (with
replacement) from the original sample. For each set, we estimate the
Milky Way parameters corresponding to the
grid cell with the maximum score as outlined above.  It has been shown
(see \cite{bf96} and references therein), that the distribution of
these bootstrapped estimates around the original one will closely
follow the intrinsic distribution of errors due to sampling.
Figure \ref{bootfig} shows the dispersion $\sigma _w$ (defined as
$\sigma _w=\sqrt{\langle w^2\rangle -\langle w\rangle ^2}$, where the
angles denote averaging) of our bootstrapped distribution of estimates
for $v_{{\rm circ}}, q$ and $p$ as a function of $N$. 
The line in bold corresponds to $\sigma
_w \propto 1/10\sqrt{N}$.  The figure shows that the circular velocity and
axis ratios of the Milky Way can be recovered to within a few 
percent using just 100
stars associated with one of the dwarf spheroidal satellites.

\section{Conclusion: You Can Judge a Galaxy by Its Tail}

In this {\it Letter}, we have explored the use of SIM measurements of
stars in tidal streams as a probe of the Galactic potential.  We find
that with five-dimensional phase-space information for only 100 stars,
we can determine the circular velocity and shape of the Galactic halo
with accuracies of a few percent.  This measurement would represent
more than an order-of-magnitude improvement in our knowledge about the
Galaxy's mass distribution (Kochanek 1996; Zhao 1998).

It should be noted, however, that our discussion has been limited to 
constraining four parameters in a specific assumed form for
the Galactic potential. 
Ultimately, as with any parameterized inversion algorithm,
the uncertainty of the method will increase with the number 
of parameters varied.
One issue which we have not considered is the degree to which the
method discussed here can be used to constrain the total extent of the
Galactic halo, in a manner analogous to that which has been applied
recently to tidal tails in merging galaxies ({\it e.g.} Dubinski et
al. 1996, 1998; Springel \& White 1998).  
Naively, we expect the measurements proposed here to
be most sensitive to the mass of the Galaxy enclosed within the region
occupied by the debris comprising a stream.  However, the unusually
high precision offered by SIM suggests that the influence of the mass
distribution on larger scales could be probed if the halo is highly
flattened.

The spatial distribution of matter in the Galaxy provides insight into
both the formation history of our Galaxy and the nature of dark
matter.  
In future work we will look at the evolution of tidal debris
in lumpy or time-dependent potentials and discuss using tidal streams
to measure not only the current state of the Milky Way, but also to
infer its history.

\acknowledgements
We thank Mike Irwin and Ed Totten for telling us about the
results of their carbon star survey
and Tim de Zeeuw and John Bahcall
for helpful comments on the paper. 
KVJ acknowledges the support of funds from the Institute for Advanced Study 
and thanks the Institute of Astronomy (Cambridge)
for the visitors grant used while putting the final touches on the paper.
HSZ would like to acknowledge a travel grant from the Leids
Kerkhoven-Bosscha Fonds and to thank 
Princeton Observatory for hospitality
during the visit that resulted in this collaboration.
DNS's work on this project was partially supported
by NASA ATP grant NAG 5-7066,
and
LH was supported in part by NASA theory grant NAG5-3059
and by the NSF under grants ASC 93-18185 and ACI-96-19019.

\clearpage

\end{document}